\documentclass[12pt,preprint]{aastex}

\def\lesssim{\mathrel{\hbox{\rlap{\hbox{\lower4pt\hbox{$\sim$}}}\hbox{$<$}}}}
\def\gtrsim{\mathrel{\hbox{\rlap{\hbox{\lower4pt\hbox{$\sim$}}}\hbox{$>$}}}}

\slugcomment{Submitted to ApJ 11/20/2003}
\shorttitle{Kerr Accretion Flows}
\shortauthors{Hirose, Krolik, De Villiers, Hawley}
\usepackage{graphicx}

\begin{document}

\title{Magnetically Driven Accretion Flows in the Kerr Metric
II: Structure of the Magnetic Field}


\author{Shigenobu Hirose, Julian H. Krolik}
\affil{Physics and Astronomy Department\\
Johns Hopkins University\\ 
Baltimore, MD 21218}
\and
\author{Jean-Pierre De Villiers, John F. Hawley}
\affil{Astronomy Department\\
University of Virginia\\ 
P.O. Box 3818, University Station\\
Charlottesville, VA 22903-0818}

\email{shirose@pha.jhu.edu; jhk@pha.jhu.edu; jd5v@virginia.edu;
jh8h@virginia.edu}

\begin{abstract}

We present a detailed analysis of the magnetic field structure found in
a set of four general relativistic 3D MHD simulations of accreting tori
in the Kerr metric with different black hole spins. Among the
properties analyzed are the field strength as a function of position
and black hole spin, the shapes of field lines, the degree to which
they connect different regions, and their degree of tangling. Strong
magnetic field is found toward small radii, and field strength
increases with black hole spin.  In the main disk body, inner torus,
and corona the field is primarily toroidal. Most field lines passing
through a given radius in these regions wander through a narrow radial
range, suggesting an overall tightly-wound spiral structure. In
the main disk body and inner torus sharp field line bends on small spatial
scales are superimposed on the spirals, but the field lines are much
smoother in the corona. The magnetic field in the plunging region is
also comparatively smooth, being stretched out radially by the
infalling gas. The magnetic field in the axial funnel resembles a split
monopole, but with evidence of frame dragging of the
field lines near the poles of the black hole.

We investigate prior speculations about the structure of the
magnetic fields and discuss how frequently certain configurations
are seen in the simulations.  For example,
coronal loops are very rare and field lines connecting high latitudes
on the event horizon to the disk are not found at all.
Almost the entire system is matter-dominated; the only force-free
regions are in the axial funnel.
We also analyze the distribution of current density, with a view toward
identifying possible locations for magnetic energy dissipation. Regions
of high current density are concentrated toward the inner torus and
plunging region. Dissipation inside the marginally stable orbit may
provide a new source of energy for radiation, supplementing the
dissipation associated with torques in the stably-orbiting disk body.

\end{abstract}


\keywords{Black holes - magnetohydrodynamics - instabilities - stars:accretion}

\section{Introduction}

Accretion onto black holes is potentially the most efficient form
of energy generation possible; it is expected that as much as
several tens of percent of the rest-mass energy of the accreted matter can
become available for radiation (Novikov \& Thorne 1973; but see also
Gammie 1999 and Agol \& Krolik 2000).  The rate and character of accretion are
regulated primarily by angular momentum transport, making the nature of
accretion torques a fundamental problem of astrophysics.  These torques
result from magnetohydrodynamic (MHD) turbulence, driven by the
magneto-rotational instability (MRI) (see the review by Balbus \&
Hawley 1998).

Large-scale numerical simulation is the best tool we have for
exploring turbulent systems.  The first attempts to describe the global
structure of accretion onto black holes employed codes solving the
equations of pseudo-Newtonian physics, i.e. Newtonian dynamics but
with a gravitational potential of the Paczy\'nski-Wiita form (Hawley
2000; Hawley \& Krolik 2001, 2002; Armitage et al. 2001; Armitage \&
Reynolds 2003; Machida
\& Matsumoto 2003).  For obvious reasons, these could provide only an
approximate description of the dynamics, and could not 
explore the effects of black hole rotation.

With the construction of a fully general-relativistic three
dimensional MHD code (De Villiers \& Hawley 2003, hereafter DH03),
it is now possible to achieve a physical description of black hole
accretion that is much more complete.  Using this code we have
simulated accretion flows onto black holes with four different spin
parameters, $a/M = 0$, $0.5$, $0.9$, and $0.998$.  These runs were
labeled KD0, KDI, KDP, and KDE in the first paper of this series (De
Villiers, Hawley, \& Krolik 2003, hereafter Paper I), which presented
both a detailed description of how these simulations were done and an
introduction to the several distinct physical regimes found within the
accretion flows.  In this paper we continue our report on these simulations
with a discussion
 of the structure of the magnetic field.

Before presenting this report, we briefly review the nature and conduct
of the simulations.  We adopt geometrodynamic units $G =
c = 1$ (Misner, Thorne, \& Wheeler, 1973), and express evolution times
in terms of black hole mass $M$.   We also follow the convention that
Greek indices represent spacetime vector and tensor components, whereas
Roman indices represent purely spatial components.
The simulations analyzed in this paper were evolved in Boyer-Lindquist
coordinates ($t,r,\phi,\theta$) on a $192 \times 64 \times 192$ spatial
grid ($r \times \phi \times \theta$). This grid is described as
``high-resolution" in Paper I. The radial grid runs from a point just
outside the event horizon (the event horizon and the inner boundary
vary with $a/M$) to $r=120\,M$ using a hyperbolic cosine distribution
to concentrate zones near the inner boundary.  The polar angle $\theta$
runs from $0.045\,\pi$ to $0.955\,\pi$ with reflecting boundaries at
the polar axis.  Only the quarter plane from $0 \leq \phi \leq \pi/2$
was included, with periodic boundary conditions in $\phi$.  The
temporal step size, $\Delta t$, is determined by the extremal light
crossing time for a zone on the spatial grid, and remains constant for
the entire evolution, as described in DH03. Step sizes are on the order
of ${10}^{-2}\,M$ for each of the four models discussed here.


The state of the relativistic fluid at each point in the spacetime is
described by its density, $\rho$, specific internal energy, $\epsilon$,
$4$-velocity, $U^\mu$, and isotropic pressure, $P$.  The relativistic
enthalpy is $h=1 + \epsilon + P/\rho$.  The pressure is related to
$\rho$ and $\epsilon$ through the equation of state of an ideal gas,
$P=\rho\,\epsilon\,(\Gamma-1)$, where $\Gamma$ is the adiabatic
exponent.  For these simulations we take $\Gamma=5/3$.  The magnetic
field is described by two sets of variables.  One is the constrained transport
(CT) magnetic field, ${\cal B}^i = [ijk] F_{jk}$, where $[ijk]$ is the
completely anti-symmetric symbol and $F_{jk}$ are the spatial
components of the electromagnetic field strength tensor. We also define
the magnetic field $4$-vector $\sqrt{4\pi}\,b^\mu =
{}^{*}F^{\mu\nu}\,U_\nu$, where $^{*}F^{\mu\nu}$ is the dual of the
electromagnetic field tensor.  Because the CT field is guaranteed to be
divergence-free, it is the magnetic field description most readily
identified with field lines.  On the other hand, the magnetic field
four-vector is the one most readily interpreted as a magnetic field for
dynamical purposes because it arises naturally in algebraically
simple forms for the stress tensor, $T^{\mu\nu}$.  The magnetic field
scalar is $||b||^2 = b^\mu\,b_\mu$, and appears in the definition of
the total four momentum, $S_\mu = (\rho\,h\ + {\|b\|}^2)\,W\,U_\mu$ ,
where $W$ is the boost factor. Magnetic pressure is given by
$P_{mag}={1\over 2}{\|b\|}^2$.  We also define auxiliary density and energy
functions $D = \rho\,W$ and $E = D\,\epsilon$, and transport velocity
$V^i = U^i/U^t$.

For each of the four simulations, the initial conditions were basically
the same:  a torus with a near-Keplerian initial angular momentum
distribution and pressure maximum at $r = 25\,M$, containing an initial
magnetic field consisting of weak poloidal field loops lying along
isodensity contours inside the torus.  Each simulation was run to time
$8100\,M$, which is approximately 10 orbits at the initial pressure
maximum.  The radial boundaries permit outflow only, so the simulations
have only a finite reservoir of matter to accrete.  However, even in
the case with the greatest integrated accretion ($a/M = 0$), only 14\%
of the initial mass was accreted in the course of the simulation.
After initial transients, the regions of the accretion flow inside
the initial pressure maximum evolve into a moderately thick, nearly-Keplerian,
highly-turbulent disk.  This structure gives the appearance of rough
statistical time-stationarity but also evolves on long timescales.

As outlined in Paper I, the quasi-steady state system is usefully
divided into five regions:  the main body of the disk, the coronal
envelope, the inner torus and plunging region, the funnel wall jet, and
the evacuated axial funnel.  In this paper we make a more extensive
examination of the properties and structure of the magnetic fields in
these regions.  Note that outside $r \simeq 25M$ (the location of the
initial pressure maximum), we are not simulating an ``accretion disk"
at all, as material in that region moves outward as it absorbs the angular
momentum taken from matter at smaller radii.  For this reason, our analysis
concentrates on the region $r \lesssim 25M$, and especially
on behavior in the vicinity of the innermost stable circular orbit.
In \S2 we examine the magnetic field strength and
distribution throughout the flow.  In \S3 we quantify properties of
the field geometry, and in \S4 we determine where field dissipation
would likely be important.  Section 5 summarizes our findings.

\section{Distribution of Magnetic Pressure}


The late-time structure of the magnetic scalar $||b||^2$ (whose physical
identification is relatively simple in the context of the fluid frame--it
is twice the magnetic energy density there) and its distribution
in the accretion flow are found in Figure~\ref{b2polview}, which shows
color contours of the azimuthal-average, $\langle {1 \over 2}
\|b\|^2\rangle$ at $t=8080\,M$ for all four simulations.  The magnetic
pressure is expressed in units of the initial maximum gas pressure in
each simulation  to permit comparison across models.  (The initial gas
pressure maximum increases with increasing black hole spin.) Gas
pressure contours (at $t=8080\,M$) are shown as white lines in order to
locate the inner torus (left panels) and the main disk body (right
panels).  The gradual thickening of the inner torus with increasing
black hole spin is highlighted by the gas pressure contours.  The left
panels show that the magnetic pressure is greatest on the top and
bottom surfaces of the inner torus, and that the regions of greatest
magnetic pressure overlie regions where the gas pressure abruptly
drops. In the funnel region, the magnetic and gas pressure profiles are
very nearly circular, though magnetic pressure is dominant in this
region. Although the magnetic pressure distribution looks irregular,
the \textit{total} pressure, gas plus magnetic, is relatively smooth;
total pressure is predominantly cylindrical in distribution outside the
funnel and spherical within (see fig.~6 of Paper I).
Figure~\ref{b2polview} shows that, near the black hole, magnetic
pressure tends to increase with black hole spin. Near $r = 5\,M$, the typical
magnetic pressure grows by roughly an order of magnitude from $a/M = 0$
to $0.998$; near the inner radial boundary, the magnetic pressure
increases by close to two orders of magnitude from zero to maximal
spin. This stands in contrast to the magnetic pressure in the main disk
body at large radii ($r > 30\,M$), which does not vary much with spin.
Figure~\ref{b2polview} also shows that magnetic pressure diminishes
rapidly both with height above the disk surface and with increasing
radius.  Gradients in magnetic pressure are very large: in just a few
gravitational radii outside the inner boundary, the pressure falls by
1--2 orders of magnitude. The magnetic pressure averaged over radial
shells falls roughly as $r^{-3}$ from the inner radial boundary to the
outer.

The spatial relation between gas and magnetic pressures is shown in
Figure~\ref{beta}, which presents the ratio of azimuthally-averaged gas
pressure to magnetic pressure, $\beta \equiv \langle P \rangle/\langle
{1 \over 2} \|b\|^2\rangle$ (see also fig.~8 in Paper I).  The main
disk body and inner torus show up quite clearly in this figure since
$\beta$ is greater than unity everywhere in these regions.  
In the dark red regions of the disk body $\beta$ can be much 
higher, as high as
$\sim 1000$ in some locations, although it is generally lower, with a
mass-weighted shell-average value of $\beta \simeq 10$--100 at all
radii within the main disk body. In the inner torus and plunging
region, $\beta$ tends to smaller values with decreasing radius for the
$a/M=0$, 0.5, and 0.9 simulations, but remains elevated in the
$a/M=0.998$ simulation, whose plunging region is both very compressed
radially in Boyer-Lindquist coordinates and not well-resolved in our
simulation. For
all but the $a/M=0.998$ simulation, the mass-weighted mean that is
10--100 in the disk body drops to $\simeq 2$--6 near the inner radial
boundary, while the volume-weighted mean $\beta$ falls from $\sim
O(10)$ near $r_{ms}$ to 0.3--1 at the inner boundary.  This decrease is
not evenly distributed, however. Through the inner torus to the
plunging region, the density contours focus toward the equatorial plane
and contours of elevated $\beta$ tend to follow. Although the aspect
ratio ($H/r$) of gas pressure-dominated material narrows in the
plunging region, in the equatorial plane itself $\beta$ remains larger
than unity all the way to the inner radial boundary.
Figure~\ref{b2polview} showed a relationship between magnetic pressure
near the black hole and the spin of the black hole.  Figure~\ref{beta}
shows that there is no comparable trend in $\beta$. Rather, as the
black hole spins faster, the gas pressure and density in the inner disk
become larger relative to the mass stored in the outer disk, and the
magnetic pressure grows proportionately.

In the coronal envelope $\beta$ is of order unity; the funnel wall jet
and other outflows within the corona have $\beta > 1$.  Generally,
$\beta$ decreases with increasing distance away from the midplane.
More than two density scale-heights above the plane, $\beta \sim 0.1$
in most places.  Because of field buoyancy one expects the corona to be
significantly magnetized.  The local stratified simulations of
Miller \& Stone (2000) found $\beta \sim 10^{-2}$ in the corona, but in
their simulation there was relatively little mass outflow, due to the
isothermal equation of state and the local disk approximation.  In our
simulations there is considerable outflow from the inner, hot portion
of the disk into the corona, which helps to maintain $\beta$ at more
modest levels.  The exception, naturally enough, is the axial funnel
which is magnetically dominated and has $\beta \ll 1$.  The centrifugal
barrier effectively prevents any significant flow of disk material into
this region. The funnel does, however, contain regions of anomalously
elevated $\beta$, due to shock heating of the tenuous funnel gas, which
tends to enhance the pressure of the gas while its density remains
extremely low. The final region of interest, the funnel-wall jet, is a
region defined by two properties: the gas is unbound ($h\,U_t < -1$)
and outbound ($S_r > 0$) (see Paper I).  Although the jet contours are
not explicitly shown in Figure~\ref{beta} (see figs.~3 and 10 of Paper
I), they occupy a narrow band with $\beta \simeq 0.3$ that lies between
the substantially lower values typical of the axial funnel and the
higher values of the corona; the jet lies poleward of the extended
high-$\beta$ radial streaks that stand out in the right panels. 
More analysis of the funnel-wall jet and the lower density outflow
within the axial funnel will be provided in a subsequent paper in this
series.

Another measure of the relative importance of the magnetic field is
given in Figure~\ref{forcefree} which shows $\langle {1 \over 2}
\|b\|^2/\rho\,h\rangle$, the azimuthally-averaged point-wise ratio of
magnetic pressure to enthalpy.  When the ratio is much greater than
unity, the gas's inertia is of minor importance to the dynamics; this
state is often called ``force-free" (e.g., Blandford 2002).  This
figure clearly distinguishes the axial funnel, the main disk body, and
the coronal envelope.  It is clear that nowhere in the main disk body,
corona, inner torus, or plunging region is the force-free condition met.
In the main disk body and most of the inner torus, the ratio is very small,
typically $\sim 10^{-4}$ in the main disk body, rising to $\sim
10^{-2}$ in the inner torus and plunging region. The ratio ranges from
$10^{-1}$ to 1 in the outer layers of the inflow through the plunging
region (compare the prediction in Krolik 1999 that $\|b\|^2/\rho \sim
0.1$--1 in the most relativistic portion of the plunging region).  In
the axial funnel, the only region where the force-free condition is
met, the ratio ranges from $\sim 100$ to 1000.

It should be noted that the GRMHD code was designed primarily for
regions where the matter inertia is nonnegligible.  The code employs a
density and energy floor that allows a dynamic range of about seven
orders of magnitude in these variables. The code also places a ceiling
on the Lorentz factor $W$ ($W_{ceiling} = 4$).  In practice these
limits come into play only at certain times within the axial funnel,
which renders the accuracy of the dynamical properties within the
funnel somewhat uncertain.  For this reason we focus mainly on the
general features of the field strength and topology in the funnel, and
give less weight to detailed calculations of energy or energy flux
there.

Azimuthally-averaged representations do not convey the entire picture.
In Figure~\ref{b2rms} we plot the fractional rms fluctuation level in
magnetic energy density relative to its azimuthal mean in the $a/M =
0.9$ simulation at $t=8080\,M$.  The rms fluctuation is defined for a
quantity $f$ by (Hawley \& Krolik 2001)
\begin{equation}\label{fluctuation}
 {\delta f \over f}\left(R,\theta\right) =
 {1 \over \langle f\rangle_\phi}  \left\{{1 \over 2\pi} \int \, d\phi \,
             \left[f - \langle f \rangle_\phi \right]^2
 \right\}^{1/2} ,
\end{equation}
where ${\langle f \rangle}_\phi$ refers to the azimuthal average at
fixed $R$ and $\theta$ of the quantity $f$.  Fluctuations in field
strength can be very large, and large fluctuations clearly delineate
the main disk body and inner torus.  In the $a/M = 0.9$ simulation the 
fractional rms fluctuation level in magnetic energy density relative to its 
azimuthal mean ranges from 0.8--2.0 within the disk.  On the other hand, both 
the plunging region and the corona are far more regular: the fractional rms
fluctuations in both zones drop to typically 0.1--0.3.  These
characteristic fluctuation levels are independent of black hole
spin.

Figures~\ref{b2polview}---\ref{b2rms} show clearly that the different
regions of the accretion flow have quite different magnetic
characteristics.  The funnel is magnetically dominated, and nearly
force-free.  In contrast, the main disk body and inner torus are gas
pressure dominated, but their evolution is nevertheless controlled by
the Lorentz forces in the MHD turbulence.  The inner edge of the disk
is a region where the inflow becomes increasingly magnetized.  The
corona is mixed; while on average the magnetic and gas pressures are
comparable, the magnetic pressure has a larger effective scale height,
and there are distinct regions where one or the other is dominant.
Nowhere, however, does the corona approach the force-free limit.

\section{Field Geometry}

Magnetic fields have directions, of course, in addition to magnitudes,
and the directions may be visualized in terms of field lines.  The
overall geometry of the magnetic field is important, since field lines
can link together different regions of the flow in a way that is not
possible by hydrodynamics alone. A convenient way to summarize the
possibilities is in terms of a modification of the field line
categorization introduced by Blandford (2002), who listed seven generic
varieties of field lines in the vicinity of an accreting black hole.
We adopt this list, but modify some of the definitions and include two
additional types (see Fig.~\ref{fieldlines}), as follows:
\begin{enumerate} \item tangled field internal to main disk, \item
loops connecting different parts of the disk through corona, \item open
field lines emerging  from the main disk, \item field lines connecting
the gas inside the plunging region to the main disk, \item high
latitude field loops connecting the black hole to the main disk, \item
open field lines from the black hole out through the funnel, \item open
field lines from the plunging region to an outflow, \item largely
toroidal field filling the corona, \item field lines connecting the
black hole to the disk through the plunging region.  \end{enumerate}

We expect that the presence or absence of some of these field line
types will be largely independent of the specific assumptions of a
given simulation.  For example, twisted field loops within the disk
body (type 1) are the generic result of the MRI.  Similarly, if there
is accretion from the disk body into the plunging region, it is very
hard to avoid creating field lines linking those two regions (type 4);
the only issue is how far they extend and the field intensity
associated with them.  Toroidal coronal field (type 8) also appears
hard to avoid whenever field is ejected into the corona.  On the other
hand, field line configurations such as type 2 may be more problematic.
Such extended field loops have been suggested as a possible origin of
high-energy emission from accretion disks.  By analogy with events in
the Solar corona, it is thought that the footpoints of magnetic
field lines in the corona may be twisted in opposite directions by orbital
shear, leading to reconnection (e.g., Blandford 2002) and a flare.  But
do such field configurations arise naturally in a turbulent disk?

Field line configurations such as types 5, 6, and 9 are of
particular interest as they offer the possibility of extracting
rotational energy from the black hole.  Whether or not there are field
lines connecting the accretion flow or the black hole's event horizon
to infinity bears directly on the issue of jet formation.  The answers
could depend on specific conditions of the accretion flow.  For
example, in these simulations we supposed that all field lines in the
initial state were closed within the initial torus of matter.  Thus,
any field lines linking the disk to infinity must be created as a
consequence of an outflowing wind or jet.  The existence of winds and
jets, and where they originate may, however, be very sensitive to the
initial conditions and the detailed thermodynamics of the accretion
flow.  Another possibility is that accretion flows drag in field lines
from large radius, thereby maintaining a connection between the
accreting matter and infinity.  Whether flows of this type exist is a
matter of conjecture and debate (e.g., Bisnovatyi-Kogan \& Ruzmaikin
1976; Thorne et al. 1986; Lubow et al. 1994).

In this section we examine the overall geometry and global connectivity
of the magnetic field in these simulations.  We begin by displaying the
general shape and character of the field lines.  Next we attempt to
describe quantitatively the degree to which field links different
regions of the flow, as well as the tangling of the field at small
scales.  We organize the discussion according to the five flow regimes
identified in Paper I and the generic field topologies given in
Figure~\ref{fieldlines}.

\subsection{Field Line Shapes}

To generate field line plots, we compute integral curves for which the
tangent vector at each point on the curve is parallel to the local
magnetic field.  In many circumstances it might be desirable to
describe the field lines in the fluid frame (using the $b^\mu$ code
variables) because in some respects that is the most physical way to
view them (see DH03 for a detailed discussion of the fluid-frame
variables).  However, the most direct way to trace these lines is to
use the CT-variables, ${\cal B}^i$, in the Boyer-Lindquist coordinates
(i.e., work directly in the coordinate frame). The CT variables are
guaranteed to be divergence-free, and have a direct relationship to the
electromagnetic field strength tensor (DH03).  In view of these
considerations, we define the local
tangent vector, \begin{equation}\label{field_line.1}
{d x^i \over d \lambda} \equiv {\cal B}^i,
\end{equation}
where $x^i$ are position vector components on the field line in the 
coordinate frame and $\lambda$ is some parametrization of the curve.
Taking the norm of the magnetic field,
\begin{equation}\label{field_line.2}
\|{\cal{B}}\|^2 = \gamma_{ij}\,{\cal B}^i\,{\cal B}^j
\end{equation}
where $\gamma_{ij}$ is the spatial sub-metric, we can rewrite this 
expression using (\ref{field_line.1}) to obtain
\begin{equation}\label{field_line.3}
\|{\cal{B}}\|^2 = \gamma_{ij}{\,d x^i\,d x^j \over d \lambda^2}
\end{equation}
This allows us to normalize (\ref{field_line.1}) to the local proper
length of the vector field,
\begin{equation}\label{field_line.4}
{d x^i \over d s} \equiv {{\cal B}^i \over \|{\cal{B}}\|}
\end{equation}
where $ds = \sqrt{\gamma_{ij}\,d x^i\,d x^j }$.  This set of
first-order differential equations is integrated with a
Runge-Kutta-Gill method using a discrete step length ($ds \rightarrow
\Delta s$) that is kept fixed ($\Delta s = 0.01\,M)$.  The integration
for each field line is begun at a starting point
$(r_o,\theta_o,\phi_o)$ and continued for a distance $s^\prime = r_o$
to either side of the starting point, where $s^\prime$ is the nominal
flat-space length, i.e., $(ds^\prime)^2 = dx^i \eta_{ij} dx^j$.
In principle, a fourth-order
Runge-Kutta scheme has an error that scales as $(\Delta s)^5$.
However, in this case, because evaluations of the right hand side of
the differential equation depend on interpolations in gridded data, the
error scales as $(\Delta s)^3$.  The fractional accumulated error in
the field line's coordinates after an integration spanning $s^\prime =
r_o$ should then be $\sim 10^{-5}(r_o/M)^{-5/2} (l/r_o)^{-3}$, where
$l$ is the physical bending scale.  To keep this error to less than one
grid cell requires that $l/r_o \lesssim 0.1 (r_o/M)^{-1}$, hence the
conservative choice in step size $\Delta s$.  To visualize these field
lines, we take the output of the integrator, which consists of a list
of grid-space coordinates for each field line, and render these using a
coordinate mapping from the Boyer-Lindquist ($r,\theta,\phi)$ triple
onto a 3D visualization grid.

We begin our description of the field geometry with the main body of the
disk, which consists, as anticipated, of turbulent, tangled toroidal
fields (type 1).  In the main disk body, the field geometry is
controlled by two effects:  orbital shear and turbulence.  The former
continually draws out radial field into azimuthal, while the latter
twists the field in all directions on scales smaller than the local
disk thickness.  The resulting field line shapes are clearly shown in
Figure~\ref{KDPfieldlines}.  For the most part, the field lines wrap around
the disk azimuthally as tightly-wound spirals, but there are numerous small
twists and tangles. The magnetic field geometry is qualitatively the
same for all black hole spins.

In the corona, the field is generally much more regular, and at times
is almost purely azimuthal.  Field-line category
8 prevails here.  There are almost
no tangles; adjacent field lines run very nearly parallel to one
another.  We must qualify this statement by acknowledging that we can
examine only selected timesteps from a full simulation and that there
is variation with time.   For example, there are some sampled time
steps in which the field in the corona more nearly resembles the
field in the main disk body as shown in Figure~\ref{KDPfieldlines},
presumably due to the passage
of magnetic pressure bubbles through the corona, a feature clearly seen
in animations of the density variable, $\rho$.  More generally,
however, poloidal field loops (type 2) appear only very rarely in the corona.
When they do, it is near the surface of the disk, where small scale
turbulent field structures emerge from the disk body.  In the disk body, the
MRI maintains turbulence, and the energy density of turbulent fluid motions
is comparable to or greater than the energy density in magnetic field.  In
the corona, on the other hand, the energy density of the magnetic field
is generally greater than the energy density in random fluid motions,
so the field adopts a shape primarily governed by equilibrating
magnetic forces, subject to the constraint imposed by the large-scale
orbital shear.

The inner part of the disk consists of the inner torus (where gas
pressure reaches a local maximum) and the plunging region.  The
plunging region lies inside the turbulence edge (Krolik \& Hawley
2002), that is, the point where the net inflow speed becomes greater
than the amplitude of the turbulent motions rather than the other way
around.  In the plunging region, the field structure is intermediate in
character between the disk body and the corona.  Because inflow starts to
dominate over turbulent fluctuations, the field is primarily controlled by
relatively smooth stretching of field lines.  Frame-dragging also helps in
this regard.  However, complete inflow dominance
is not achieved until the flow reaches deep into the plunging region.
In the $a/M=0.9$ simulation, the innermost stable circular orbit is at
$2.32\,M$, at least a small amount of tangling persists throughout the
plunging region, while in the $a/M =0.998$ 
simulation (not shown), where the plunging region is poorly resolved (Paper I),
extensive tangling remains down to the inner boundary.  Field-line types 4
and 9 predominate in this region, as the plunging region is linked magnetically
to both the main body of the disk and the inner boundary of the simulation.

     The magnetic field in the axial funnel is created
in the initial accretion event when relatively strong field lines first
reach the horizon.  Material drains off the field lines and they
rapidly expand out from the equator to fill the funnel.  Once
established, these field lines remain largely unchanged through the
remainder of the simulation, and in polar force balance with the
corona.  These are field lines of type 6,
a split-monopole configuration with loosely-wound helices.  The
dominant component is radial, but there is an azimuthal component that
increases with black hole spin and proximity to the black hole.  The
specific angular momentum of material in this region is 2--3 orders of
magnitude smaller than in the nearby disk, so nearly all the rotation
is due to frame-dragging.  Conversely, at
increasing distances from the hole, the funnel field lines become
almost purely radial.  Because the angular momentum barrier prevents any matter
located initially in the torus from finding its way into the axial
funnel, there are almost no field line connections linking any part of
the disk to the outflow.  There are likewise no field lines that start
at high-latitude points on the inner radial boundary, pass through the
axial funnel, and lead to the disk body.  That is, field lines of types 5
and 7 are entirely absent.

The field lines in the funnel-wall jets (not shown) resemble those of
the axial funnel, except that they are more tightly wound, considerably
more so for the high-spin simulations, $a/M=0.9$ and 0.998. This is
consistent with the observations made in Paper I that the gas in the
funnel-wall jets has specific angular momentum comparable to the value
at the injection points, the surface of the inner torus.

The effect of frame dragging on the field lines is especially prominent
at the smallest radii, just outside the event horizon.
Figure~\ref{framedrag} makes this point dramatically.  When $a/M = 0$,
field lines threading the event horizon outside the equatorial plane
are almost perfectly radial.  Those passing through the horizon in the
plane are swept back azimuthally because they are being carried in with
matter having substantial angular momentum.  However, as $a/M$
increases, the tightness of the azimuthal winding of all field lines,
whether associated with accreting matter or not, grows sharply more
dramatic as frame-dragging enforces rotation in the Boyer-Lindquist
frame.

Close inspection of the $a/M = 0$ panel in Figure~\ref{framedrag} reveals
a curious effect, specific to that case.  As the inner boundary is approached,
the lapse causes the transport speeds to fall, reversing the sense of shear.
Because the stream lines in the equatorial plane are more tightly-wound
than the field lines, the shear controls the correlation between ${\cal B}^r$
and ${\cal B}^\phi$, and the field lines bend backward.

\subsection{Field Line Connectivity}

An important property of magnetic field lines is their ability to link
distant regions through magnetic tension.  Indeed, it is the ability of
the field to connect orbiting fluid elements that is the cause of the
MRI and the resulting angular momentum transport in accretion flows.
In the context of Kerr black holes, magnetic fields offer the additional
possibility of linking the rotating spacetime near the hole with more
distant regions, potentially extracting rotational energy from the hole
itself.

In an attempt to quantify the length scales connected by the field
lines, we define the \textit{wandering index}, a simple measure of the
range of coordinate values spanned by a given field line.
Specifically, we define the wandering index $\Delta X$ for a poloidal
coordinate $X = (r/r_o,\theta)$ as the difference between the maximum
and minimum $X$-value along a particular field line of length $s^{\prime} =
r_o$.  Next, to compute
mean values of this index we choose approximately 100 field line
starting points within a specified range of angular coordinates on the
spherical shell with radius $r_o$ at each of ten different times
spanning the last $720\,M$ of the simulation.  The probability density
for the starting points is the density of field-lines piercing that radial
shell; i.e., it is proportional to $(\sqrt{g_{rr}}/\sqrt{-g}){\cal B}^r$.  In
these calculations, we do not take account of field lines that reach the
inner boundary.

The wandering index for different coordinates takes on characteristic
values in certain limits.  For example, when the field line is purely
toroidal, both $\Delta r/r_o$ and $\Delta\theta$ are identically zero.
Purely radial field has $\Delta r/r_o = 1$ and $\Delta \theta = 0$.
Field lines confined to the equatorial plane have $\Delta \theta = 0$,
and the size of their $\Delta r/r_o$ index indicates the tightness of
their spiral winding: $\Delta r/r_o \rightarrow  0$ is the
tight-winding limit, and $\Delta r/r_o = 1$ is the purely radial
limit.  We caution, however, that the scaling of $\Delta X$ with
integration path-length $s^{\prime}$ is difficult to estimate.

In the disk body, a typical field line connects regions separated by
$\Delta r/ r_o \simeq 0.2$, with little dependence on $a/M$
(Figure~\ref{wander}).  Thus, on this scale of field line length,
the field lines are moderately wound (as seen in
Figure~\ref{KDPfieldlines}).  For the $a/M=0$ and 0.5 models, the
radial distance spanned increases inside the marginally stable orbit as
the flow changes from one that is primarily turbulent to one that is
primarily spiral inflow.  For the high-spin models ($a/M =
0.9$ and 0.998), inside the static limit ($r_{static}=2\,M$ in the
equatorial plane, independent of spin) frame-dragging ensures that most
of the field line's physical length is stretched in azimuth rather than
radius, so that $\Delta r/r_o$ falls.

The wandering index for the polar direction is also shown in
Figure~\ref{wander}.  In the disk body, field lines move typically
0.08--0.15 radians in polar angle in the course of traversing one
radius in length.   By contrast, the typical aspect ratio of these
disks as found from the gas density scale-height is $H/r \simeq 0.2$
(Paper I).  Thus, although the magnetic field scale-height is
roughly double that of the gas, the vertical distance across which an
individual field line moves in a length of one radius is less than a
single gas scale-height.  The focusing of the inflow toward the
equatorial plane further reduces the polar wandering index as $r$
decreases.

\subsection{Field Line Tangling}

The preceding measures mostly reflect the larger-scale properties of
the field line geometry.  While a field line may not extend over
significant distances, and hence may have a small wandering index, it may
nevertheless be far from smooth and uniform.  To quantify the
small-scale structure of the field lines, we define a field line
\textit{tangling index}, $\Psi$, 
by accumulating the angle between adjacent tiny
segments along field lines of length $s^{\prime} = r_o$:  
\begin{equation}\label{tangledef} 
\Psi \equiv \sum_k \arccos \left(\eta_{ij}\,n_{(k+1)}^i\,n_{(k)}^j\right),
\end{equation} 
where $n_{(k)}^i$ is the unit three-vector in the
direction from point $k-1$ to point $k$ along the field line and
$\eta_{ij}$ is the flat space metric.  For this index, the sequence of
points is {\it not} the set of points generated at each integration
step but rather a set along the integrated field line separated by
$\Delta s^{\prime} = 0.025\,r$.  We chose this scheme because the
radial coordinate grids of the simulations were very nearly
logarithmic, with separations varying between $\Delta r = 0.025\,r$ for
the $a/M= 0$ case to $\Delta r = 0.028\,r$ for the $a/M = 0.998$ run.
By choosing this sampling scheme, we assure ourselves of achieving a
fixed resolution level across the different regions of the
simulations.

In the limit that the resolution scale is small compared to the length
scale of large-angle field line bending, the tangling index has a
simple interpretation.  For small individual angles, (\ref{tangledef})
becomes \begin{equation} \Psi \simeq \sum_k 2 \left[|1 -
\eta_{ij}\,n_{(k+1)}^i\,n_{(k)}^j |\right]^{1/2}.  \end{equation} When
the individual bend angles are small, $n_{(k)}^i$ is effectively a
continuous function of distance along the field line and so may be
expanded as a Taylor function in $s^{\prime}$.  Taken to second-order,
\begin{equation} 
\Psi \simeq \sum_k \sqrt{2} \left[\eta_{ij}\,n^i (s_{(k)}^{\prime})
	 {d^2 n^j \over d{s^{\prime}}^2}\right]^{1/2} \Delta s^{\prime}
	 = \sum_k {1 \over R(s_{(k)}^{\prime})}\,\Delta s^{\prime},
\end{equation} 
where $R(s_{(k)}^{\prime})$ is the local radius of
curvature.  Thus, in the limit $\Delta s^{\prime} \rightarrow 0$,
\begin{equation} \Psi \rightarrow \int \, {1 \over
R(s^{\prime})}\,ds^{\prime} .  \end{equation} That is, the tangling
index is the ratio of the length of the field line to the harmonic mean
of its radius of curvature. In the unlikely limit of a perfectly
straight field line, $\Psi = 0$. For a circular segment in any plane,
$\Psi=1$ (including the equatorial plane, where such a segment would be
purely toroidal).  Because we compute a discrete approximation to the
integral with 40 (i.e. $1/0.025$) individual angles, the maximum
possible tangling index is $\Psi = 40\,\pi \simeq 125$.

The results of computing this measure are displayed in
Figures~\ref{tangle} and \ref{tangledist}.  Over one radius of field
line length, the accumulated small-scale bends in the disk body sum to
4 -- 7 radians, with a tendency for the tangling to increase with
increasing $a/M$.  In approximate terms, the distribution of individual
bend-angles $dN/d\sigma$ inside the disk is proportional to
$\sigma^{-2}$, so the contribution to the total bending from
each decade of angular scale is roughly constant, and the cumulative
bending distribution rises logarithmically with bend angle
(Fig.~\ref{tangledist}).  Such a result might have been expected given
the turbulence in the disk. Inside the marginally stable orbit, the
degree of small-scale tangling drops sharply as turbulence gives way to
a comparatively smooth flow. The deeper inside the plunging region the
flow reaches, the less tangled the field lines. In the coronal region
the field lines become much smoother, with the tangling index falling
to 1--2.  In the limit of a perfectly smooth azimuthal field line, the
index would be identically 1 as described above; thus, the coronal
field lines are very smooth indeed.

\section{Other Field Properties}

\subsection{Symmetry of ${\cal B}^{\phi}$}

Since the initial condition consists of purely poloidal field loops
centered on the equator, the initial radial field is antisymmetric
across the equator, as is the sign of ${\cal B}^\phi$ that results
from subsequent shear.  This holds true through much of the evolution.
During the first two orbits the total toroidal flux grows rapidly and
antisymmetrically.  However, the accretion flow that forms is
unstable to vertical oscillations across the equator, breaking the
perfect antisymmetry.  As a result, the {\it net} toroidal flux
in each hemisphere gradually declines as matter initially in one hemisphere
is mixed into the other.  The global net toroidal flux can depart from
zero once field begins to enter the black hole, but it never becomes
as large as the net flux in one or the other hemisphere.
At the end of the simulations there are still some obvious antisymmetries
in ${\cal B}^\phi$ in the corona, and near the funnel in the $a/M=0.9$ and
0.998 simulations, but no systematic antisymmetry remains within the disk
itself.  At the end of the simulations, the net ${\cal B}^\phi$ flux
compared to the total, i.e., $\int {\cal B}^\phi \sqrt{\gamma} \, d^3 x^i/
\int |{\cal B}^\phi | \sqrt{\gamma} \, d^3 x^i$ is $\simeq 1\%$ when
$a/M =0$ and 0.998, 10\% when $a/M = 0.5$, and 8\% when $a/M = 0.9$.

\subsection{Correlation with velocity}

In computing idealized axisymmetric, smooth, and time-steady accretion
solutions with large-scale magnetic fields, Li (2003a,b) shows that if
the ratio $(b^\phi u^r)/(b^r u^\phi) = 1$, MHD accretion onto black
holes can occur with zero electromagnetic transport of energy.
Presumably this conclusion is closely linked to the fact that when this
ratio is unity, the velocity and magnetic fields in the disk plane are
parallel.  In the ideal MHD limit, where $\vec E = \vec v \times \vec
B$, this means $\vec E = 0$.  When that is so, the Poynting vector must
also be identically zero, and no energy is transported
electromagnetically in the course of accretion.  We can explicitly
measure this ratio with data from our simulations.

As illustrated in Figure~\ref{liratio}, in general this ratio is quite
different from unity.  In most of the corona, disk body, inner torus, and
plunging region it falls in the range $-0.3$ -- $+0.3$, but there are
locations near the funnel wall where the
ratio spans a considerably wider range, with thin regions in the
vicinity of the jet having values near unity. Values near unity are
also found at the surface of the plunging flow above and below the
main equatorial inflow, but even there the ratio fluctuates
substantially in space and in time.  We therefore conclude there is
nothing to prevent significant electromagnetic energy fluxes in these
accretion flows.

\subsection{Dissipation Regions}

These simulations were conducted with no explicit resistivity, and,
consequently, they do not address where the magnetic field is
dissipated or at what rate.  However, regions of high current density
are {\it candidates} for regions of high magnetic dissipation because
high current density may trigger anomalous resistivity through mechanisms
such as ion-acoustic turbulence (e.g., as suggested for the Solar corona
by Rosner et al. 1978).  We can locate
these candidate dissipation regions by computing the current density
from our simulation data. The current 4-vector is given by 
\begin{equation}
J^\mu = { 1 \over 4\,\pi}\,\nabla_\mu\,F^{\mu \nu},
\end{equation}
where $\nabla_\mu$ is the covariant derivative and $F^{\mu \nu}$ the
electromagnetic field-strength tensor. The covariant derivative 
simplifies to a simple derivative using the anti-symmetry of $F^{\mu \nu}$
(eq. 23 of DH03)
\begin{equation}
J^\mu = { 1 \over 4\,\pi\,\sqrt{-g}}\,\partial_\mu\left(
 \alpha\,\sqrt{\gamma}\,g^{\mu \lambda}\,g^{\nu \xi}\,F_{\lambda \xi}
 \right),
\end{equation}
where $F_{\lambda \xi}$ is directly related to the CT magnetic field
and EMFs (eqs. 14 and 35 of DH03), the EMFs being simple
functions of the CT magnetic field and transport velocities $V^i$.
Evaluation of $J^\mu$ using code variables requires data for three
adjacent time steps, since time derivatives must be evaluated. The
current density then follows directly from these calculations, $\|J\|^2
= J^\mu\,J_\mu$.

Figure~\ref{dissipation} shows azimuthally-averaged $\|J\|^2$ at a late
time in each of the four simulations.  The values are normalized to
the initial torus total energy, $E_0$ ($ \equiv \int \, d^3 x^i \, \sqrt{-g}
\, T^t_t$ in the initial state), to facilitate comparison between
simulations. In all cases, regions of high current density in the accretion
flow are found in extended sheets that run roughly parallel to density contours
in the inner torus and plunging regions; currents in the main disk body
are significantly weaker and less well organized, while the coronal
envelope is quiescent. The axial funnel shows a current distribution
similar to the distribution of magnetic pressure
(Fig.~\ref{b2polview}). In all cases, the dynamic range between
$\|J\|^2$ in one of the sheets and in the adjacent inner torus material
is often two orders of magnitude.  As with the magnetic pressure, there
are also several systematic trends with black hole spin.  The vertical
thickness of the region in the disk body where high current density is
found increases with $a/M$.  The absolute level of $\|J\|^2$ (even after
normalization to $E_0$) also increases with $a/M$, climbing by about a
factor of 300 from $a/M = 0$ to $a/M = 0.998$.

Figure~\ref{J2eq} shows $\|J\|^2/E_0$ in the equatorial plane 
($\theta = \pi/2$) at a late time in each of the four simulations. This
figure emphasizes the spiral structure of the currents.  Here again,
the steep radial gradient in current density can be seen.  The regions
of most intense current are almost all found within $\simeq 2 r_{ms}$.

If we knew how to relate current density to dissipation rate, these
maps would provide a basis for a map of heating in the accretion
flow.  In the absence of a physical model for the relation between
current density and dissipation, we can instead give a qualitative
sense for the heating distribution by integrating $\|J\|^2$ over volume
in the disk, i.e., computing ${\cal J} \equiv \int \, d\theta d\phi \,
\sqrt{-g} \, \|J\|^2$, but excluding those regions with
$\rho < 10^{-4}\rho_{\rm max}$, where $\rho_{\rm max}$ is the greatest
density found anywhere in the simulation at that time.  We ignore low-density
regions in order to concentrate on possible heating in the accretion flow
proper.  The results are shown in Figure~\ref{integdiss},
again normalized to $E_0$. The left panel shows the integrated current
density over the entire range of radii; the right panel shows the same
quantity plotted as a function of $r/r_{ms}$. Both panels show the
systematic increase in current density with black hole spin, particularly
for $r<10M$. In the disk body, ${\cal J}$ declines steeply with increasing
radius.  Between $r_{ms}$ and $10r_{ms}$, $-2 < d\ln {\cal J}/d\ln r < -1$,
and the slope steepens at still larger radii.  In the $a/M=0$ simulation, the
integrated current density levels off slightly through the marginally stable
orbit before rising sharply again near the inner boundary. For the
other three simulations, the integrated current density has a sharp
break upward near or inside $r_{ms}$, rising one to two orders of magnitude
as the inner radial boundary is approached. Since frame dragging has been
shown to increase magnetic pressure and azimuthal stretching of the
field lines, this dramatic increase is not surprising. If the current
density is indicative of the dissipation density, much of the total
dissipation, especially around the most rapidly-spinning black holes,
could occur very deep in the relativistic potential.

\section{Discussion and Summary}


In Paper I we divided the accretion flow into five distinct regions,
each with its own characteristics, as follows: the main disk body, the
inner torus and plunging region, the coronal envelope, the funnel, and
the funnel wall jet.  In this paper we have examined the magnetic field
strengths and topologies in these regions.

\subsection{Field Characteristics in the Different Flow Regions}

The main disk body is characterized by MHD turbulence due to the MRI.
As has been shown in previous studies, the magnetic field
is very tightly wound by the differential rotation, but with
significant poloidal tangling due to turbulence.  The field
wandering index shows that the poloidal field lines extend farther in
the radial direction than in the polar.  The field energies are
subthermal, with plasma $\beta \approx 10$--100 on average.  Strong
fluctuations are the rule, however, and $\beta$ can exceed
1000 in some portions of the main disk body.

In the plunging region, the flow transitions from turbulence-dominated
to spiraling inflow.  A consequence of this is that the tight field
winding loosens somewhat as the radial component of the velocity grows
relative to the azimuthal component.  This behavior is quantified
approximately by the field-wandering and field-tangling indices, which show
(see Figs.~\ref{wander} and \ref{tangle}) that there is a sharp
change of behavior near the marginally stable orbit.  Field lines
inside that radius travel farther in radius and much less in polar
angle than in the disk body.  This is a consequence of the transition
from turbulent motions to more regular spiral inflow; that is, this
break is yet another way of marking the turbulence edge defined by
Krolik \& Hawley (2002).  Inside the plunging region the field is
amplified relative to the fluid pressure.  Near the horizon above and
below the equator, $\beta$ falls from $\sim 10$ near $r_{ms}$ to less
than 1. The equatorial flow itself maintains $\beta \sim 10$ or greater
throughout the plunging region.  These conclusions apply most clearly to the
$a/M=0$, 0.5, and 0.9 simulations; the plunging region in the
$a/M=0.998$ simulation is less well-resolved.

The field in the corona is, on average, in equipartition with the
thermal energy there, and is dominated by the toroidal
component; field line plots look smooth compared to those in the
disk body.  There are relatively extended poloidal fields, but these
are more associated with outflowing toroidal coils.  When field
emerges from the disk body it carries some of the turbulent tangling
with it, but this seems to be rapidly smoothed out.  Gas is carried
into the corona as well as magnetic field, so the corona does not
become a magnetically dominated force-free region.  The magnetic
field does have a larger scale-height than the gas pressure,
though.

By contrast, the field in the axial funnel is essentially radial
outside the ergosphere, with a toroidal component that becomes larger
with increasing black hole spin.  In the funnel the magnetic field
energy dominates over the matter energy, and the field can be regarded
as force-free.  The magnetic energy is not large compared to the magnetic
energy in the remainder of the flow, however.  The magnetic pressure in the
funnel is in equilibrium with the total pressure in the corona and inner disk.
From the field-connectivity data, we see that the magnetic field
structure may be divided into two almost wholly independent regions:
the accretion flow proper and the axial funnel.  Field lines passing
through the funnel almost never enter the other regions.

In the funnel-wall jet, the magnetic field configuration resembles that
of the axial funnel, but the field lines are more tightly wound,
consistent with the observation that the gas in the jet has specific
angular momentum comparable to that at its injection point, above the
inner torus. The jet is also less strongly magnetized than the axial
funnel, with $\beta \simeq 0.3$.

The categorization of field line types listed in Table~\ref{fieldlinetable}
and illustrated in Figure~\ref{fieldlines} provides a handy summary of
the global field structure.  It is important to recognize that
only a subset of all possible field line types are found with any frequency in
these simulations; the table indicates which these are.  One of the most
important global properties of the field is its inter-region connectivity.
From this point of view, the different regions fall into two groups:
the main disk body is linked to both the corona
and the inner torus, while many field lines stretch from the inner
torus to the plunging region.  On the other hand, the field lines in
the axial funnel have essentially no connection to the other regions.

\subsection{Astrophysical Implications}

\begin{figure}[ht]
  \epsscale{1.0}
  \plotone{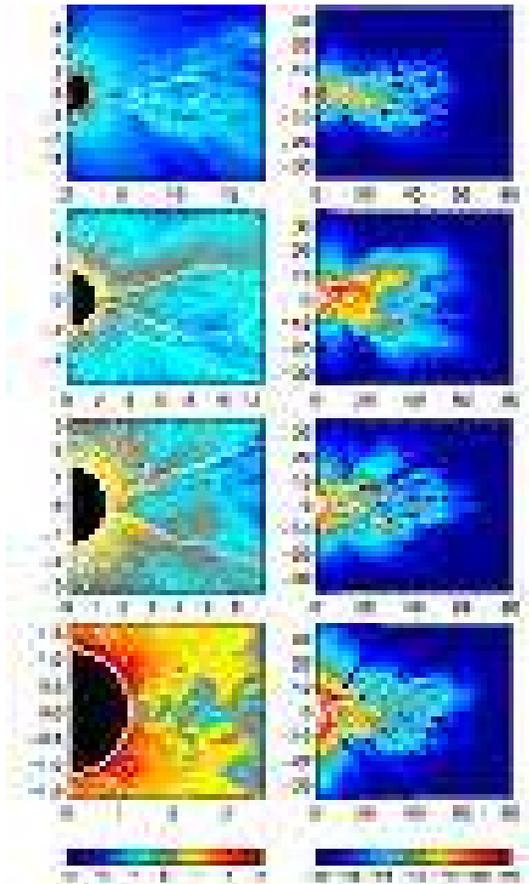} 
  \caption{\label{b2polview} The azimuthally-averaged magnetic pressure, 
  $\langle {1 \over 2} \|b\|^2\rangle$, at $t = 8080 \,M$ normalized to the 
  maximum initial gas pressure, $P_{max}$, in simulations with 
  $a/M = 0$ (first row from top), 0.5 (second row), 0.9 (third row), and 
  0.998 (fourth row). The left column shows the region inside $r = 3\,r_{ms}$,
  plotted on a logarithmic scale that emphasizes the increasing dominance
  of magnetic pressure with black hole spin in the innermost region of the
  computational domain. The right column shows the main body of the disk,
  in a color scale that emphasizes the dominance of gas pressure
  in the disk body (the upper color ranges are saturated here). Three
  azimuthally-averaged gas pressure contours are also shown,
  $\langle P_{gas}\rangle$, scaled to the initial pressure maximum, 
  $0.1\,P_{max}$ (dotted line), $P_{max}$ (solid line),
  $10\,P_{max}$ (dashed line).   
  } 
\end{figure}

\begin{figure}[ht]
  \epsscale{1.0}
  \plotone{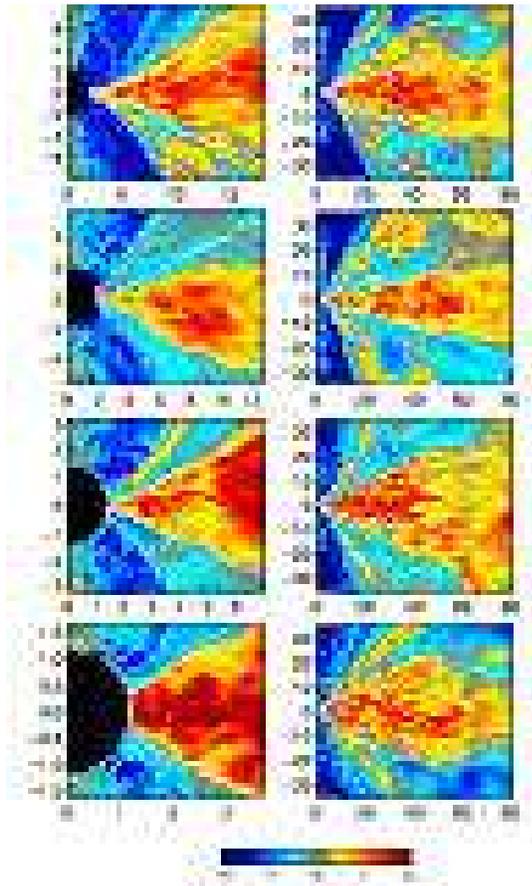} 
  \caption{\label{beta} The azimuthally-averaged ratio of gas to magnetic
  pressure, $\beta = \langle P \rangle/\langle {1 \over 2}\|b\|^2\rangle$, at 
  $t = 8080 \,M$ in simulations with $a/M = 0$ (first row from top), 
  0.5 (second row), 0.9 (third row), and 0.998 (fourth row). 
  The left column shows the region inside 
  $r = 3\,r_{ms}$, the right column shows the main disk body; all are plotted 
  on the same logarithmic scale.  Three azimuthally-averaged density contours 
  are also shown, $\langle \rho \rangle$, scaled to the initial pressure 
  maximum, $0.1\, \rho_{max}$ (solid line), $10^{-2}\, \rho_{max}$ 
  (dotted line), $10^{-3}\, \rho_{max}$ (dashed line).   
}
\end{figure}

\begin{figure}[ht]
  \epsscale{1.0}
  \plotone{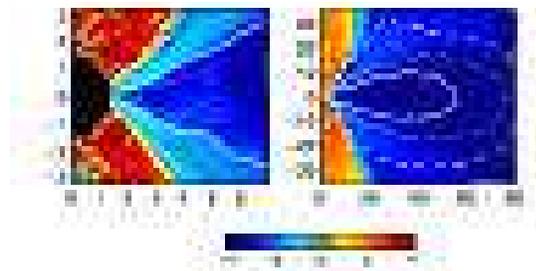}
  \caption{\label{forcefree} The azimuthally-averaged ratio of magnetic
  pressure to enthalpy, $\langle {1 \over 2}\|b\|^2 / (\rho\,h)\rangle$, at 
  $t = 8080 \,M$ for the simulation with $a/M = 0.9$. 
  The left panel shows the region inside 
  $r = 3\,r_{ms}$, the right panel shows the main disk body; both are plotted 
  on the same logarithmic scale.  Three azimuthally-averaged density contours 
  are also shown, $\langle \rho \rangle$, scaled to the initial pressure 
  maximum, $0.1\, \rho_{max}$ (solid line), $10^{-2}\, \rho_{max}$ 
  (dotted line), $10^{-3}\, \rho_{max}$ (dashed line). Images from 
  the other three simulations are very similar. 
} 
\end{figure}

\begin{figure}[ht]
  \epsscale{1.0}
  \plotone{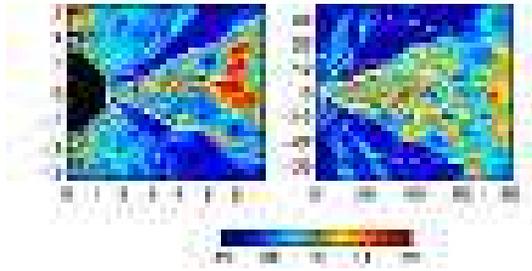}
  \caption{\label{b2rms} The fractional fluctuation in azimuth for 
  $\|b\|^2$ at $t = 8080 \,M$ in the simulation with $a/M = 0.9$.  
  The left panel shows the region inside $r = 3\,r_{ms}$, the right panel 
  shows the main disk body; both are plotted on the same linear scale. Images 
  from the other three simulations are very similar. } 
\end{figure}

\begin{figure}[ht]
    \epsscale{1.0}
    \plotone{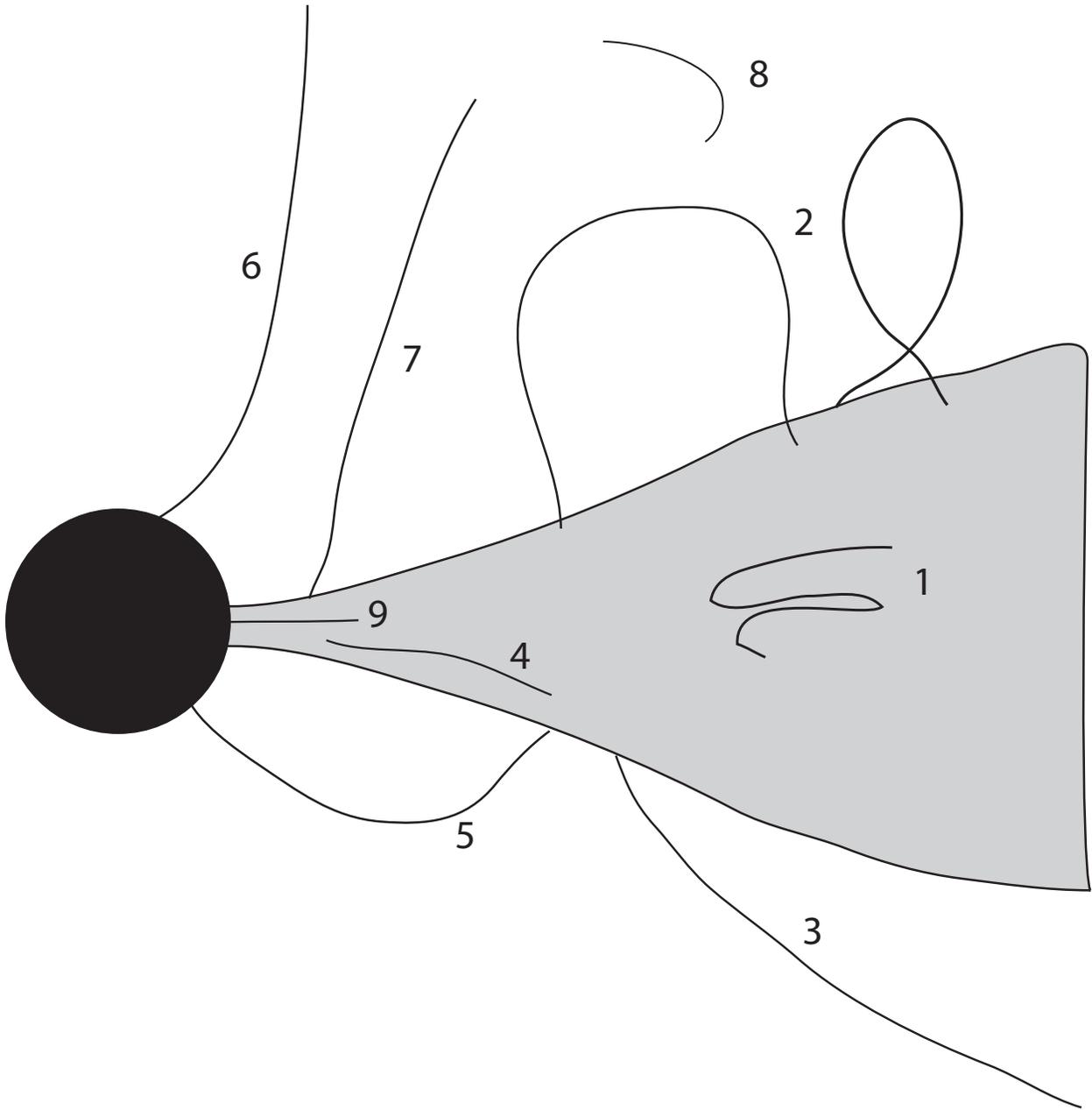} 
    \caption{\label{fieldlines} 
     Schematic diagram (after Blandford 2002) showing possible magnetic 
     linkages between different regions of the accretion flow and the black 
     hole.
     } 
\end{figure}

\begin{figure}[ht]
    \epsscale{1.0}
    \plotone{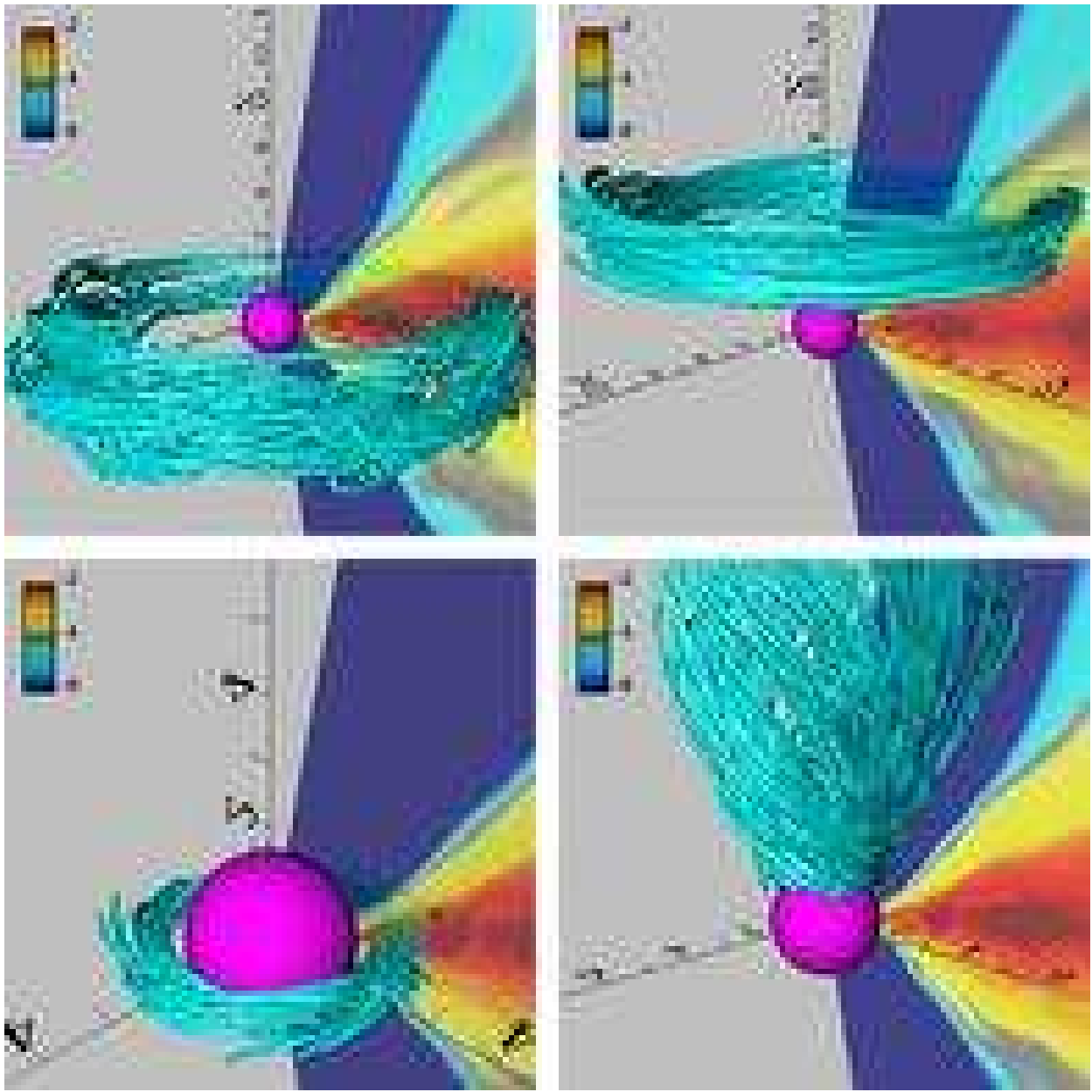}
    \caption{\label{KDPfieldlines} Sample field lines in four regions of
    the $a/M = 0.9$ simulation.  Upper left panel is the disk body (i.e.,
    $0.45\pi<\theta<0.55\pi$, $r_o=10M$); upper right is the corona
    ($0.32\pi<\theta<0.37\pi$, $r_o=10M$); lower left is the plunging region
    ($0.475\pi<\theta<0.525\pi$, $r_o=2M$); lower right is the axial funnel
    ($0.1\pi<\theta<0.2\pi$, $r_o=5M$).  All four are at $t=7760M$.  Background
    colors are contours of $\log (\rho)$, calibrated by the color-bar.}
\end{figure}

\begin{figure}[ht]
    \epsscale{1.0}
    \plotone{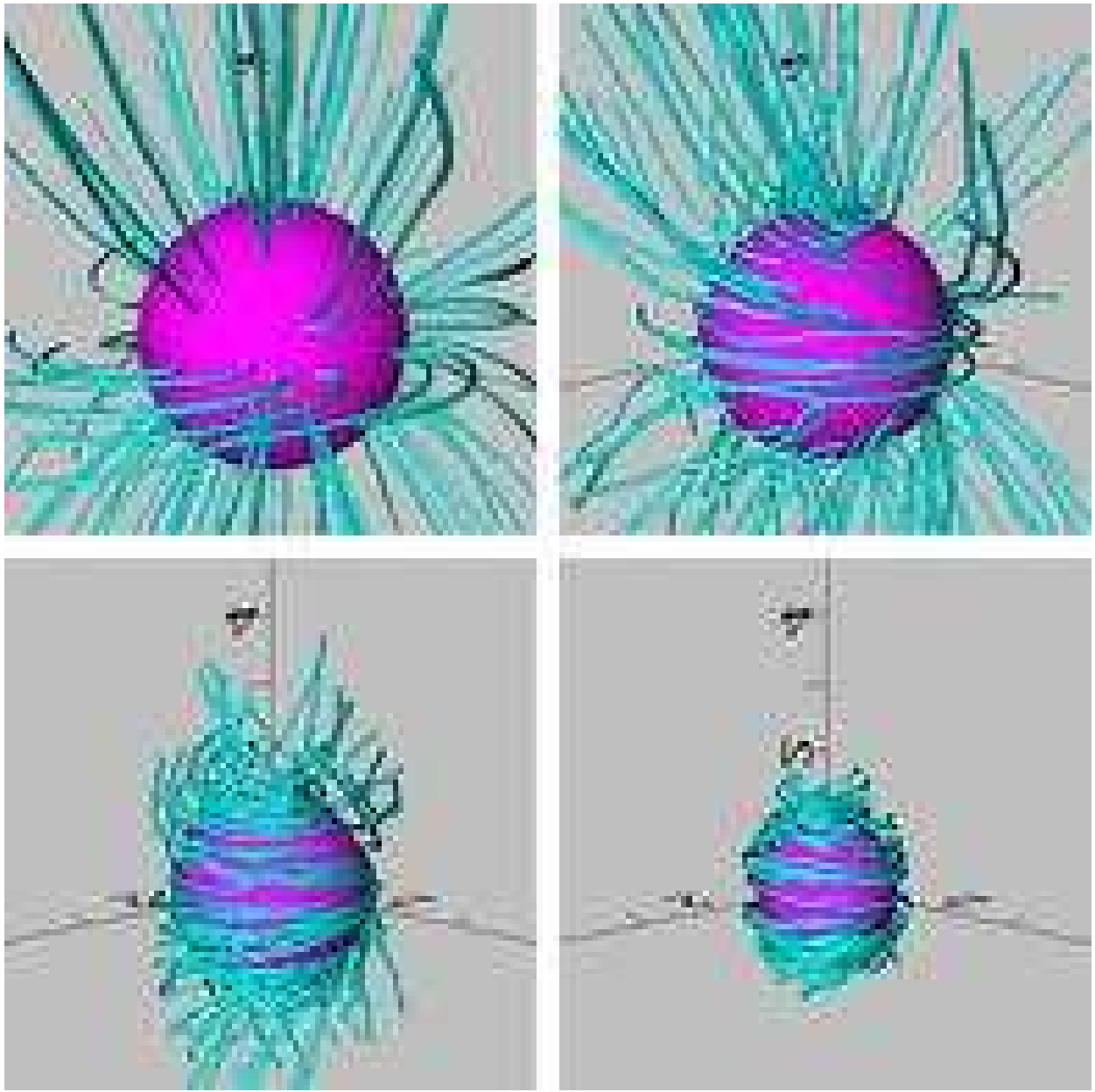} 
    \caption{\label{framedrag} Sample field lines threading the innermost
radial boundary in simulations with $a/M = 0.$, 0.5, 0.9, and 0.998.} 
\end{figure}

\begin{figure}[ht]
    \epsscale{1.0}
    \plotone{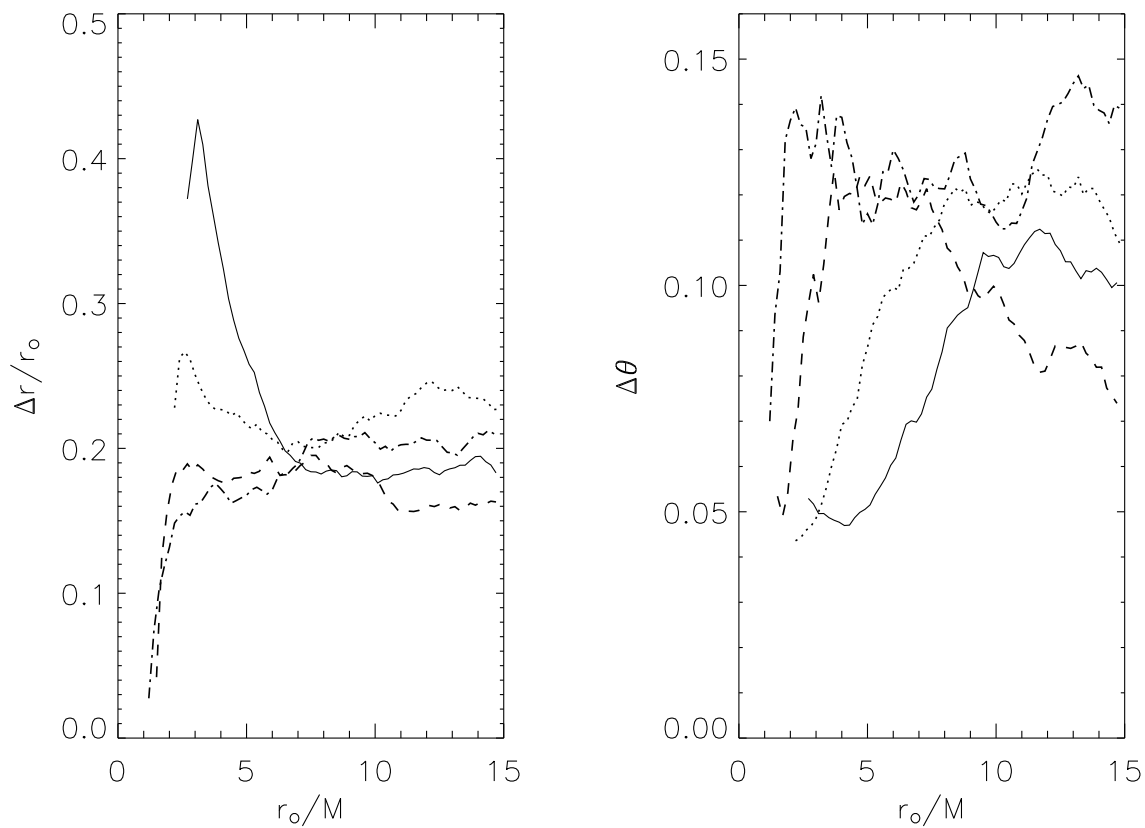} 
    \caption{\label{wander} Wandering index (see text for definition) as a
    function of radial coordinate of the field line's starting point for
    each of the four black hole spins considered. The curves are:
    $a/M=0$ simulation (solid line); $a/M = 0.5$ simulation (dotted line);
    $a/M = 0.9$ simulation (dashed line); and $a/M = 0.998$ simulation
    (dash-dot line).
    Left-hand panel: Radial wandering. 
    Right-hand panel: Polar angle wandering.}
\end{figure}

\begin{figure}[ht]
    \epsscale{1.0}
    \plotone{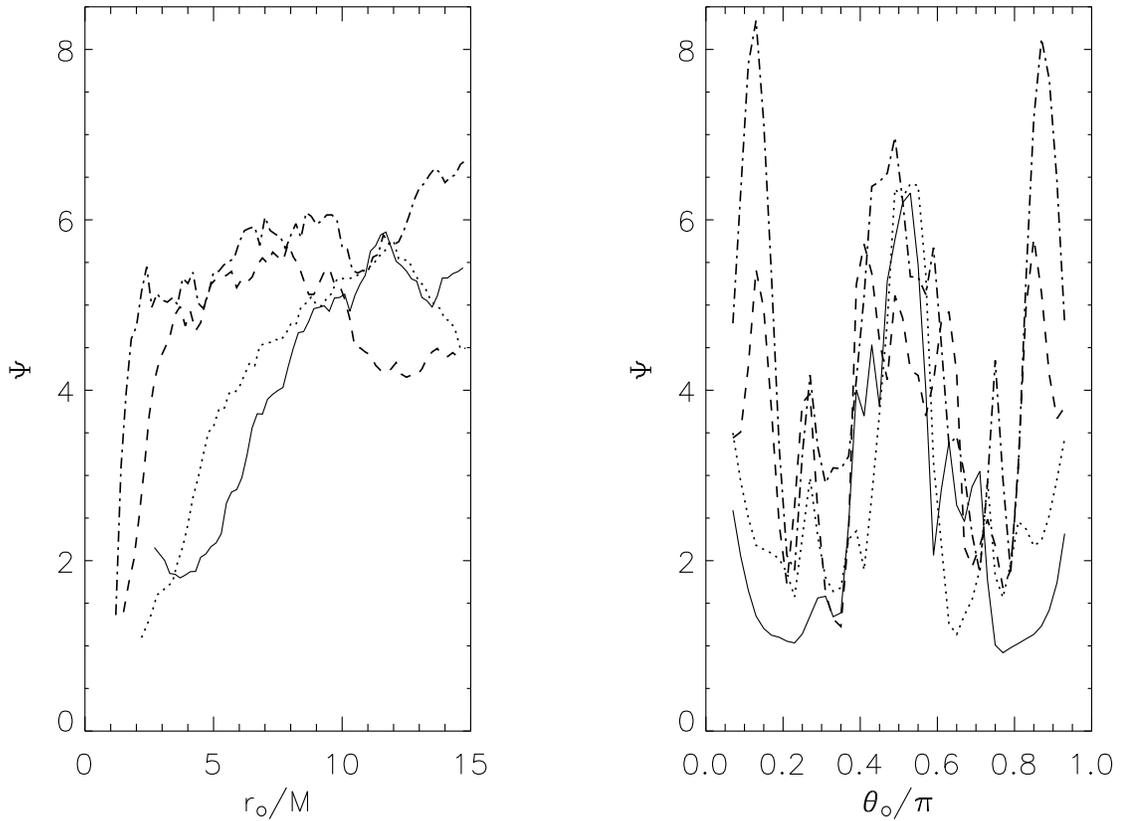} 
    \caption{\label{tangle} The field-tangling index $\Psi$ as a function of
$r_o$ inside the disk body ($0.45\pi < \theta < 0.55\pi$: left panel) and
$\theta_o$ ($10M < r < 15M$: right panel). The curves are:
    $a/M=0$ simulation (solid line); $a/M = 0.5$ simulation (dotted line);
    $a/M = 0.9$ simulation (dashed line); and $a/M = 0.998$ simulation
    (dash-dot line).} 
\end{figure}

\begin{figure}[ht]
    \epsscale{1.0}
    \plotone{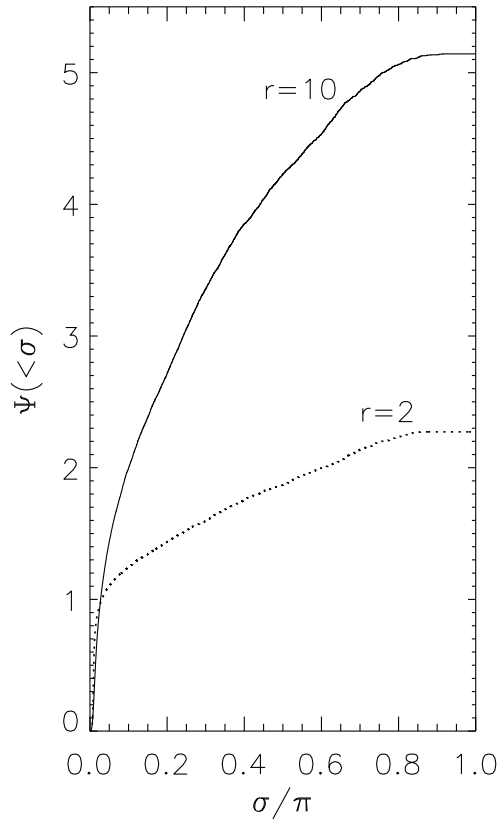} 
    \caption{\label{tangledist} The cumulative distribution of tangling
$\Psi (< \sigma) = \int_{0}^{\sigma} \, d\sigma^{\prime} \sigma^{\prime}
dN/d\sigma^{\prime}$ averaged over ten late-time snapshots in the
$a/M = 0.9$ simulation at the radii shown and within $0.05\pi$ radians
of the equatorial plane.  Here $dN/d\sigma$ is the number distribution
of bend angles.} 
\end{figure}

\begin{figure}[ht]
 \epsscale{1.0}
 \plotone{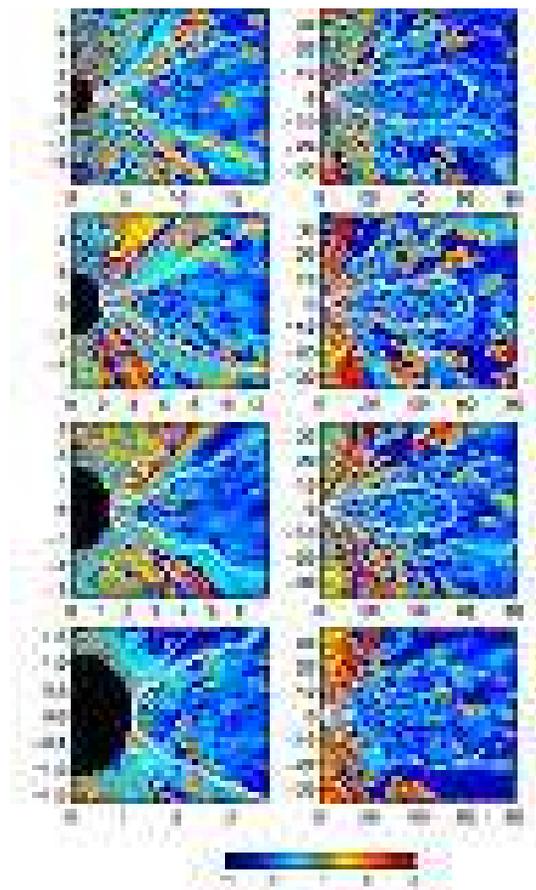} 
 \caption{\label{liratio} The azimuthally-averaged ratio 
 $\langle b^\phi\,u^r \rangle/\langle b^r\,u^\phi \rangle$ , at $t =8080 \,M$
 in simulations with $a/M = 0$ (first row from top), 0.5 (second row), 0.9 
 (third row), and 0.998 (fourth row). The left column shows the region inside 
 $r = 3\,r_{ms}$, the right column shows the main disk body; all are plotted 
 on the same linear scale.  Three azimuthally-averaged density contours 
 are also shown, $\langle \rho \rangle$, scaled to the density 
 maximum, $10^{-3}\, \rho_{max}$ (dashed line), $10^{-2}\, \rho_{max}$ 
 (dotted line), $0.1\, \rho_{max}$ (solid line). } 
\end{figure}

\begin{figure}[ht]
  \epsscale{1.0}
  \plotone{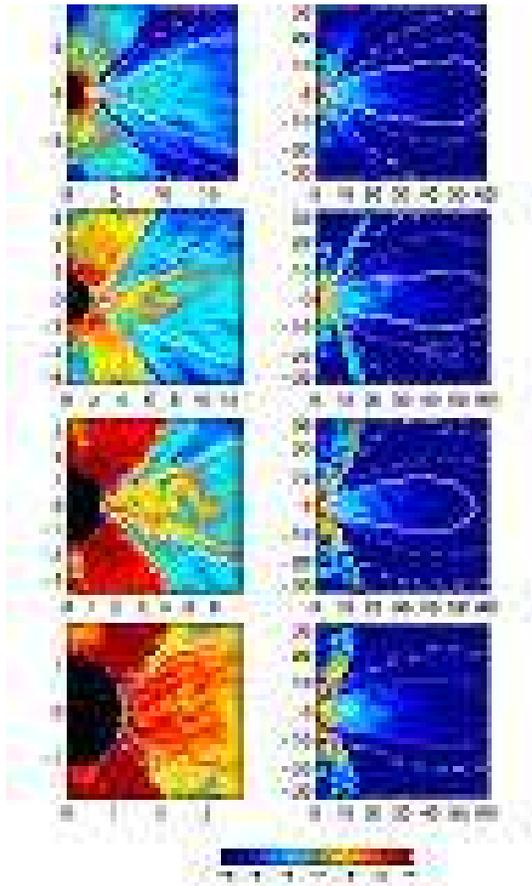} 
  \caption{\label{dissipation} Azimuthally-averaged $\|J\|^2$ at 
  $t = 8080 \,M$ in simulations with $a/M = 0$ (first row from top), 0.5 
  (second row), 0.9 (third row), and 0.998 (fourth row). 
  The left column shows the region inside 
  $r = 3\,r_{ms}$, the right column shows the main disk body; all are plotted 
  on the same logarithmic scale.  Three azimuthally-averaged density contours 
  are also shown, $\langle \rho \rangle$, scaled to the density 
  maximum, $10^{-3}\, \rho_{max}$ (dashed line), $10^{-2}\, \rho_{max}$ 
  (dotted line), $0.1\, \rho_{max}$ (solid line). } 
\end{figure}

\begin{figure}[ht]
 \epsscale{1.0}
 \plotone{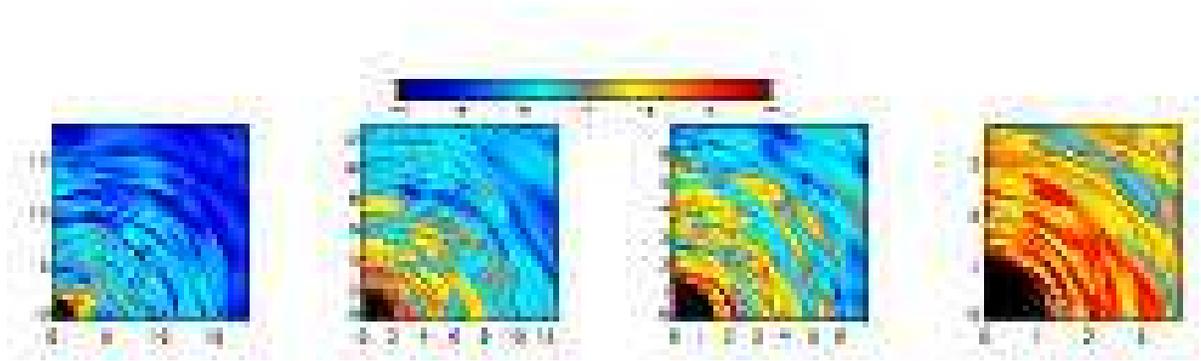} 
 \caption{\label{J2eq} $\|J\|^2$ in the equatorial plane ($\theta = \pi/2$)
 at $t = 8080 \,M$ in simulations with (from left to right) $a/M = 0$, 0.5,
 0.9, and 0.998.  In each case, the outermost radius shown is $3r_{ms}$.
 All are plotted on the same logarithmic scale as Figure~\ref{dissipation}.} 
\end{figure}

\begin{figure}[ht]
    \epsscale{1.0}
    \plotone{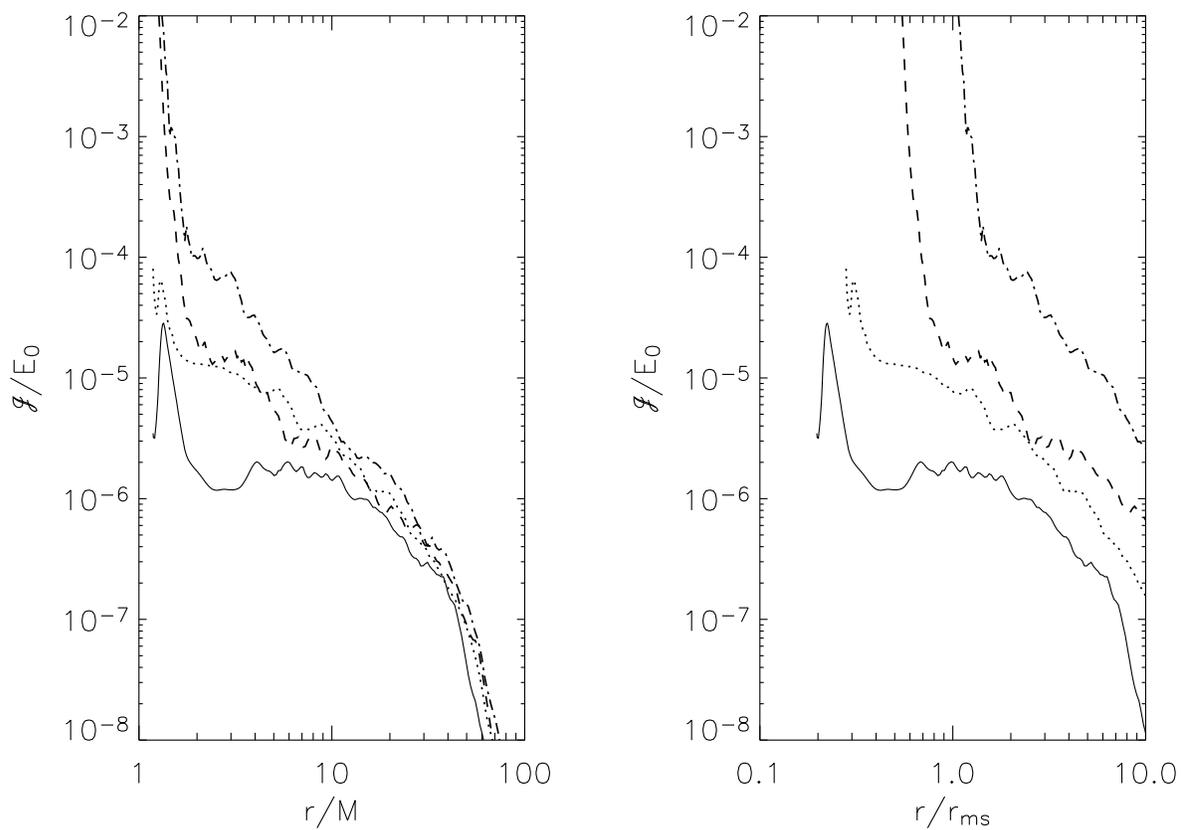} 
    \caption{\label{integdiss} Shell-integrated current density, 
    (see text for definition) plotted
    as a function of $r/M$ (left panel) and $r/r_{ms}$ (right panel) 
    for the $a/M=0$ simulation (solid line), 0.5 simulation (dotted line),
    0.9 simulation (dashed line), and 0.998 simulation (dash-dot line).
    The integral over $\theta$ includes only that region where
    $\rho > 10^{-4}\, \rho_{max}$, thereby excluding the funnel.
} 
\end{figure}

The simulation results provide some insights into potentially important
astrophysical processes that might occur in black hole accretion.  For
example, Livio et al. (1999) argued that energy extracted
electromagnetically from the accretion disk proper would always
outweigh that extracted from spinning black holes because the product
of magnetic stress and area would be larger for the disks than for
black hole event horizons.  As Figure~\ref{b2polview} shows, the total
intensity of the magnetic field increases sharply inward.  This is as
true of the poloidal portion as the toroidal.  Increases by factors
$\sim 30$ from $r=5M$ to the plunging region are typical.  Although it
is true that, outside the funnel region, there is no large-scale
poloidal field in this simulation, such large contrasts in field
strength from disk to plunging region raise doubts about the Livio et al.
hypothesis.  Moreover, the force-free field in the evacuated funnel seems best
configured to permit processes closely related to the Blandford-Znajek
mechanism (Blandford \& Znajek 1977), while the strength of the funnel field
is comparable to that in the plunging region.  These results suggest
that accretion could well lead to a strong enough field
near the event horizon that a genuine Blandford-Znajek process might
yield more energy than competing processes associated more directly with
the main disk body.

It is often asserted (in fact, frequently in connection with studies of
the Blandford-Znajek mechanism) that the magnetospheres of black holes
should be, in large part, effectively force-free (e.g., Okamoto 1992,
Ghosh \& Abramowicz 1997, Ghosh 2000, Park 2000, Blandford 2002,
Komissarov 2002).  In our simulations we find that the only region
where the force-free approximation is applicable is the funnel.
In the disk proper and most of the corona, the field structure is
thoroughly matter-dominated; in the outer corona and parts of the
plunging region, the energy density in the magnetic field begins to
approach $\rho h$, but never exceeds it.  Thus, the relativistic MHD
approximation may be better-suited to work in this area than the
relativistic electrodynamic approximation.  It is possible, though, that
cooling would alter the properties of the corona somewhat by significantly
reducing the gas pressure scale height relative to that of the magnetic
field.

A common suggestion for the origin of high-energy emission from
accretion disks around black holes is that the footpoints of magnetic
field lines in the corona are twisted in opposite directions by orbital
shear, leading to reconnection (e.g., Blandford 2002).  This process is
initiated by establishment of a radial magnetic field link between two
fluid elements arranged so that the outer one is advanced in orbital
phase relative to the inner one.  As the inner fluid element catches up
to and passes the outer, an X-point would be formed.  We see little
evidence of this sort of process.  What appears to happen instead is that
when radial magnetic field links are created, it is between two fluid
elements at nearly the same azimuthal angle (the closer two points are,
the more likely they are to be magnetically connected).  The orbital
shear then creates a long azimuthal field line as the fluid element at
the smaller radius moves ahead of the other fluid element.  If the
inner fluid element moves a full $2\pi$ in azimuth ahead of the outer
one before reconnection occurs, the field lines wrap parallel to one
another, rather than crossing near their footpoints.


It is generally believed that regions of high current density are likely
to promote anomalous resistivity, and therefore be markers for regions
of concentrated heat dissipation.  Although we find that, if anything,
current sheets are especially rare in the corona, they are very common
and very intense in the innermost parts of the accretion flow, particularly
the plunging region.
The possibility of substantial heat release inside $r_{ms}$ has an
interesting consequence.  In the Novikov-Thorne theory of relativistic
accretion disks, the energy made available for radiation is the
difference between the potential energy lost by accreting matter and
the work done by accretion stresses.  Because they assumed that all
stresses vanish inside $r_{ms}$, there is no energy released inside that
radius.  If the Novikov-Thorne model is generalized to allow non-zero
stress at $r_{ms}$ (e.g., Agol \& Krolik 2000), the amount of energy
dissipated inside the disk rises, but it is still determined in
essentially the same way.  Dissipation inside $r_{ms}$ is an entirely
new and independent mechanism of energy release.  It is not necessarily
associated with angular momentum transport; it could result from
entirely local processes (e.g., Machida \& Matsumoto 2003).  Nonetheless,
if the time to radiate this heat is shorter than an infall time, the
photons released could carry off the energy to infinity, adding to the
light seen by distant observers.

One of the main purposes of carrying out these simulations across a
broad range of black hole spins is to extract information previously
unobtainable from pseudo-Newtonian simulations of accreting tori.
The spacetime rotation caused by spinning black holes is a notable
example of this sort of effect.
Perhaps the most striking spin-dependent property of the magnetic
fields is the increase in magnetic pressure near the black hole with
increasing spin. This pressure increase is found at the surface of the
inner torus and plunging region (i.e., off the equatorial plane), as
well as deep in the axial funnel. There is also a corresponding
increase in gas pressure in this region, so that the ratio of pressures
($\beta$) remains effectively unchanged. The growth of magnetic
pressure is due in part to the dragging of field lines by the black
hole, and in part to the inward displacement of the marginally stable
orbit with black hole spin, which tends to allow the MRI to operate
closer to the event horizon. There is also a strong spin-dependent
growth in the intensity of currents in the vicinity of the black hole.
In the $a/M=0$, 0.5, and 0.9 simulations, the high current regions are
organized in sheets through the inner torus; these current
sheets also have a spiral character, inherited from gas motion in the
vicinity of the plunging region. The $a/M=0.998$ simulation shows a
much more intense and turbulent current distribution, reflecting the
fact that the plunging region is extremely narrow in Boyer-Lindquist
coordinates and the the inner
torus extends deep into the ergosphere. This strong spin-dependence of
the currents suggests that magnetic dissipation may be especially
intense in accretion flows near extreme black holes.

\acknowledgements{SH would like to thank Toshio Uchida for many helpful
suggestions and comments. JHK thanks Ethan Vishniac for frequent helpful
conversations.  This work was supported by NSF grant
AST-0070979 and PHY-0205155, and NASA grant NAG5-9266 (JPD and JFH),
and by NSF Grant AST-0205806 (JHK and SH). The simulations were carried
out by JPD on the Bluehorizon system of NPACI.}

\clearpage

\begin{deluxetable}{cll}
\tablecaption{\label{fieldlinetable}Field Line Varieties 
(after Blandford 2002)}
\tablecolumns{3}
\tablehead{\colhead{Type Number} & \colhead{Description} & \colhead{Common?}}
\startdata
1 & tangled within the disk body & yes \\
2 & loops linking regions of disk through corona & no \\
3 & poloidal from disk to infinity & no \\
4 & linking accreting matter in plunging region to disk body & yes \\
5 & linking high latitudes on event horizon to disk body & no \\
6 & poloidal from event horizon to infinity & yes \\
7 & poloidal from matter in plunging region to infinity & no \\
8 & smooth tightly-wrapped coronal spiral & yes \\
9 & poloidal field linking horizon to accretion flow & yes 
\enddata
\end{deluxetable}

\clearpage

\clearpage

\clearpage

\clearpage

\clearpage

\clearpage

\clearpage

\clearpage

\clearpage

\clearpage

\clearpage

\clearpage

\clearpage

\clearpage

\end{document}